\documentclass[preprint,showpacs,preprintnumbers,amsmath,amssymb]{revtex4}

\usepackage{graphicx}
\usepackage{dcolumn}
\usepackage{bm}
\newcommand{\be}{\begin{equation}}
\newcommand{\ee}{\end{equation}}
\newcommand{\bea}{\begin{eqnarray}}
\newcommand{\eea}{\end{eqnarray}}

\begin{document}


\title{Absorptive corrections for vector mesons. Matching to complex mass scheme and longitudinal corrections.}

\author{L. A. Jim\'enez P\'erez and G. Toledo S\' anchez}
\affiliation{Instituto de F\'{\i}sica,  Universidad Nacional Aut\'onoma de M\'exico, AP 20-364,  M\'exico D.F. 01000, M\'exico}

\date{\today}

\begin{abstract}
Unstable spin-1 particles are properly described by including absorptive corrections to the electromagnetic vertex and propagator, without breaking the electromagnetic gauge invariance. We show that the modified propagator can be set into a complex mass form, provided the mass and the width parameters, which are properly defined at the pole position, are replaced by energy dependent functions fulfilling the same requirements at the pole. We exemplify the case for the $K^*(892)$ vector meson, where the mass function deviates around 2 MeV from the $K\pi$ threshold to the pole position.
The absorptive correction depends on the mass of the particles in the loop. For vector mesons, whose main decay is into two pseudoscalar mesons ($PP'$), the flavor symmetry breaking induces a correction to the longitudinal part of the propagator. Considering the $\tau^- \to K_S\pi^-\nu_\tau$ decay, we illustrate these corrections by obtaining the modified vector and scalar form factors. The $K_S\pi^-$ spectrum is described considering  the  $K^*(892)$ and $K^{'*}(1410)$ vectors and one scalar particle. Nonetheless, for this case, the correction to the scalar form factor is found to be negligible.
 \end{abstract}


\maketitle

\section{Introduction}
 The absorptive quantum loop corrections introduce the finite width effects for spin-1 unstable particles, while keeping electromagnetic gauge invariance, based in two main observations:  In quantum field theory the width is naturally included in the imaginary part of the self-energy of the particles and, the Ward identity is respected at all orders in perturbation theory. These facts are exploited in the so-called fermion/boson loop schemes, which modify consistently the propagator and the electromagnetic vertex  \cite{baur95,argyres95,beuthe,bls}. On the other hand, the so-called  complex mass scheme \cite{complexmass} has proven to be successful to account for the unstable feature by replacing the squared mass $M^2 \to M^2+iM\Gamma$ in all the Feynman rules, where $\Gamma$ is the full decay width. It has been pointed out that there is an equivalence between both approaches \cite{bls,david}, provided a renormalization of the vector field is invoked \cite{david}. All over this statement, the mass and width parameters in the complex mass scheme are assumed to correspond to the measured ones.
In this work we show that, starting from the absorptive corrected propagator, in order to set it into the complex mass form, the mass and width must be replaced by energy dependent functions
corresponding to the width and mass at the pole position. We exemplify the case for the $K^*(892)$ vector meson, where the mass function deviates around 2 MeV from the $K\pi$ threshold to the pole position, and the width function exhibits a different behavior compared to the uncorrected energy dependent width.
Another feature that we study, arising from the absorptive corrections, is the flavor symmetry breaking effect, which induces a correction to the longitudinal part of the propagator. Considering the $\tau^- \to K_S\pi^-\nu_\tau$ decay, we exemplify this correction by obtaining the modified vector and scalar form factors for the $K\pi$ spectrum considering the $K^*(892)$ and $K^{'*}(1410)$ vectors and one scalar particle. Nonetheless, the correction to the scalar form factor is found to be negligible.

\section{Absorptive corrections to the propagator}
Let us recall some facts that give ground to the inclusion of the absorptive correction to the propagator we will use later.
The propagator for a vector particle (V) of mass $m$ and momentum $q$ at tree level, can be set as:
\begin{equation}
D_0^{\mu \nu} (q)= - \dfrac{\imath T^{\mu \nu} (q)}{q^{2} - m^{2}} + \dfrac{\imath L^{\mu \nu} (q)}{m^{2}},
\label{propwcero}
\end{equation}
where $T^{\mu \nu} (q) \equiv g^{\mu \nu} - q^{\mu} q^{\nu} /q^{2}$ and $L^{\mu \nu} (q) \equiv q^{\mu} q^{\nu} /q^{2}$, are the  transversal and longitudinal projectors respectively. The electromagnetic vertex for a charged vector  $V(q_1) \rightarrow V(q_2) \gamma(k)$  at tree level, can be set as:
\begin{equation}
\Gamma_0^{\mu \nu \lambda} = g^{\mu \nu} (q_1 + q_2)^{\lambda} - g^{\mu \lambda} (q_1 + k)^{\nu} - g^{\nu \lambda} (q_2 - k)^{\mu}.
\label{verticewwfcero}
\end{equation}
The above expressions for the vertex and propagator satisfy the Ward identity $k_\mu \Gamma_0^{\mu \nu \lambda} =[iD_0^{\nu \lambda} (q_2+k)]^{-1}- [iD_0^{\nu \lambda} (q_2)]^{-1}$.
Upon the inclusion of the contribution from the imaginary part of the one-loop corrections,  the propagator is modified in a generic form as:
\begin{equation}
D^{\mu \nu} (q) = - \dfrac{\imath T^{\mu \nu}}{q^{2} - m^{2} + \imath Im \Pi^{T} (q^{2})} + \dfrac{\imath L^{\mu \nu}}{m^{2} -\imath Im \Pi^{L} (q^{2})},
\label{punlazo}
\end{equation} 
where the absorptive contribution induced by the particles in the loop are split in a transverse and longitudinal part:
\begin{equation}
Im \Pi^{\mu\nu} (q)= Im \Pi^{T} (q^2)T^{\mu \nu}+Im \Pi^{L} (q^2)L^{\mu \nu}.
\end{equation}
Similarly, the vertex becomes
$\imath e \Gamma^{\mu \nu \lambda} = \imath e (\Gamma_0^{\mu \nu \lambda} + \Gamma_1^{\mu \nu \lambda})$,
where $\Gamma_1^{\mu \nu \lambda}$ contains the loop corrections. The Ward identity, which is fulfilled order by order, relates the loop contributions by requiring them to satisfy:
\begin{equation}
k_\mu \Gamma_1^{\mu \nu \lambda} = \imath Im \Pi^{\nu\lambda} (q^2_1) - \imath Im \Pi^{\nu\lambda} (q^2_2).
\label{niw}
\end{equation}
For the $W$ gauge boson, the scheme considers that the imaginary part of such loops are dominated by fermions, while for vector mesons like the $\rho$ and $K^{*+}$, whose main decay is into two pseudoscalar mesons ($PP'$), bosons are the natural particles in the loop.
The transverse and longitudinal part of the absorptive contribution of the self-energy have been calculated for the fermion \cite{baur95,argyres95,beuthe} and the boson \cite{bls} loops. Here, we focus in the bosonic case, whose analytic expressions are:
\begin{equation}
Im \Pi^{T} (q^{2})=\Gamma(q^2)\sqrt{q^2};   \hspace*{1cm}   \Gamma(q^2)=\frac{g^2}{48 \pi q^2}\left( \frac{\lambda(q^2,m_c^2,m_n^2)}{q^2}\right)^{3/2}
\label{width0}
\end{equation}
and
\begin{equation}
 Im \Pi^{L} (q^{2})=-\frac{g^2 \lambda^{1/2}(q^2,m_c^2,m_n^2)}{16 \pi}\left(\frac{m_c^2-m_n^2}{q^2}\right)^2.
 \label{longitudinal}
\end{equation}
where $\lambda(x,yz)\equiv x^2+y^2+z^2-2xy-2xz-2yz$.   $m_c$ and $m_n$ are the masses of the charged and neutral particle in the loop respectively and $g$ denotes the strength of the coupling of the vector meson to this pair. Thus, the imaginary correction introduces the width ($\Gamma(q^2)$) in the transversal part and a flavor symmetry breaking term in the longitudinal part of the vector propagator.\\
The real part correction will provide a mass shift which is neglected all over this description, breaking analyticity, as the finite width effect is expected to be the more relevant feature. Here we stick to the approach used to obtain the above results, but certainly a full consideration of this correction is needed.

\section{Matching to the Complex mass scheme}
The propagator in the complex mass scheme takes the following form:
\begin{equation}
D^{\mu \nu} (q) = i
\left( 
\frac{ -g^{\mu \nu}+ 
\frac{q^\mu q^\nu}{M^2-iM\Gamma}  }
{ q^{2} - M^{2} + i M\Gamma}  
 \right)
\label{pCM}
\end{equation} 
where the width can be either  constant or energy dependent. Let us work out the propagator obtained in the boson loop scheme (Eqn. \ref{punlazo}) to bring it into the above complex mass form. Replacing the transverse and longitudinal projectors, it is explicitly given by:
\begin{equation}
D^{\mu \nu}(q) = \frac{i}{q^2-m^2+i Im\Pi^T(q^2)} 
\left( 
 -g^{\mu \nu}+ \frac{q^\mu q^\nu}{m^2-i Im\Pi^L(q^2)} 
\left(\frac{ q^{2}+i Im\Pi^T(q^2)-i Im\Pi^L(q^2) }{q^2}\right)
 \right).
\label{propagadorgral}
\end{equation} 
We can split the transverse and longitudinal part in terms of  two generalized functions,  $\gamma$ and $\beta$  respectively, times the squared momentum dependence as follows:
\begin{equation}
Im \Pi^{T} (q^{2})\equiv \gamma q^2,  \hspace*{2cm} Im \Pi^{L} (q^{2})\equiv\beta q^2.
\label{fun}
\end{equation}
Using the full expression for the corrections Eqns. (\ref{width0}) and (\ref{longitudinal}), the new functions exhibit a mild dependence on $q^2$. Upon these definitions, the propagator becomes
\begin{equation}
D^{\mu \nu}(q) = \frac{i}{(q^2-m^2+i \gamma q^2)} 
\left( 
 -g^{\mu \nu}+ \frac{q^\mu q^\nu}{m^2-i \beta q^2} 
\left( 1+i (\gamma-\beta) \right)  
 \right).
\label{propagadorgral2}
\end{equation} 
This can be set in the complex mass form times a global factor, which can be absorbed into the normalization of the vector state \cite{david,scherer,renormalization}:
\begin{equation}
D^{\mu \nu} (q) = \frac{i}{(1+i(\gamma-\beta))} 
\left( 
\frac{ -g^{\mu \nu}+ 
\frac{q^\mu q^\nu}{M^2-iM\bar \Gamma}  }
{ q^{2} - M^{2} + i M\bar \Gamma}  
 \right)
\label{propagadorgralCM}
\end{equation} 
where the $M$ and $\bar \Gamma$ functions are defined by:
\begin{equation}
M^2 \equiv \frac{m^2-\beta q^2(\gamma -\beta)}{ 1+(\gamma-\beta)^2}
\label{massf} 
\end{equation}
and \begin{equation}
M \bar \Gamma \equiv \frac{\beta q^2+m^2(\gamma-\beta)}{1+(\gamma-\beta)^2}.
\label{widthf}
\end{equation}
Note the explicit momentum dependence of such functions. 
An example of the isospin limit is the correction to the $\rho$ meson, where pions are the on-shell particles allowed in the loop around the $\rho$ pole. In this case, the mass difference between the neutral and charged particle is small and the {\it longitudinal contribution can be effectively neglected} ($\beta=0$).
The correction to the  $K^*(892)$ is an example where we can not neglect the longitudinal contribution any longer, since the particles in the loop are a kaon and a pion, which have very different masses.  In cases involving vector mesons made up of heavy-light quarks, the corrections become more important,  like the $D^*(2010)$, where the two pseudoscalars are the $D$ and $\pi$  mesons, with even larger mass difference.
 Let us now show the interpretation and relation of the parameters involved to the experimental measurements.

\begin{figure}
\includegraphics{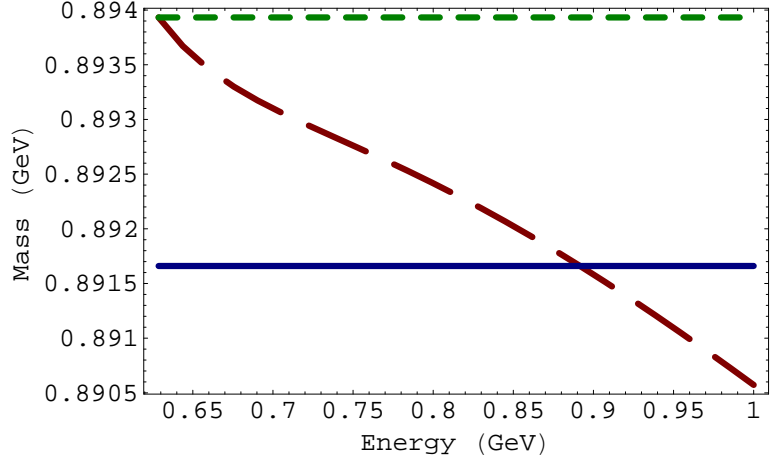}
\caption{\label{massp} Mass behavior from threshold to 1 GeV. The short-dashed line (green) is the mass at threshold, the solid line (blue) is the experimental value (PDG) and the long dashed line (red) is the mass function obtained in this work by bringing the propagator into the complex mass form.}
\end{figure}

\subsection{Mass}
Let us recall the mass function Eqn. (\ref{massf}) obtained above.
We can interpret $m$ as the value of the mass at threshold ($M_{th}^2=m^2$), by noticing that the $\beta$ and $\gamma$ functions are null at threshold ($\beta_{th}=\gamma_{th}=0$).
We can find its value by considering that a consistent definition of mass in QFT corresponds to the the pole of the propagator ($M_0^2=q^2$), which can be extracted from experimental data:
\begin{equation}
M_0^2 \equiv \frac{m^2-\beta_0 M_0^2(\gamma_0 -\beta_0)}{ 1+(\gamma_0-\beta_0)^2},
 \end{equation}
where the subindex ($_0$) corresponds to the evaluation of the functions at $q^2 =M_0^2$.
Then, 
 \begin{equation}
m^2 =( 1+(\gamma_0-\beta_0)^2 ) M_0^2 +\beta_0 M_0^2(\gamma_0 -\beta_0).
 \end{equation}
Let us illustrate the values of such parameters and functions for the case of the $K^*(892)$ meson, taking the experimental mass value ($M_0=891.66$ MeV) \cite{pdg}, we obtain:
$\beta_0=-0.03$,$\gamma_0 =0.06$ and
$m=893.93$ MeV. In Figure \ref{massp}, we exhibit the corresponding behavior of the mass function compared to the threshold and experimental value. There is a difference between the function at threshold and at the pole position of around 2 MeV, which may be important on precision measurements. This may contribute to explain the mass difference observed between the neutral and charged $K^*$ \cite{pdg}, extracted considering the complex mass form, where, the neutral particle is not influenced by these corrections and should correspond to the mass at threshold. In addition, we notice that the mass equation, once $m$ has been fixed, allows for a second pole solution below threshold, at $M=0.28$ GeV.
 
\begin{figure}
\includegraphics{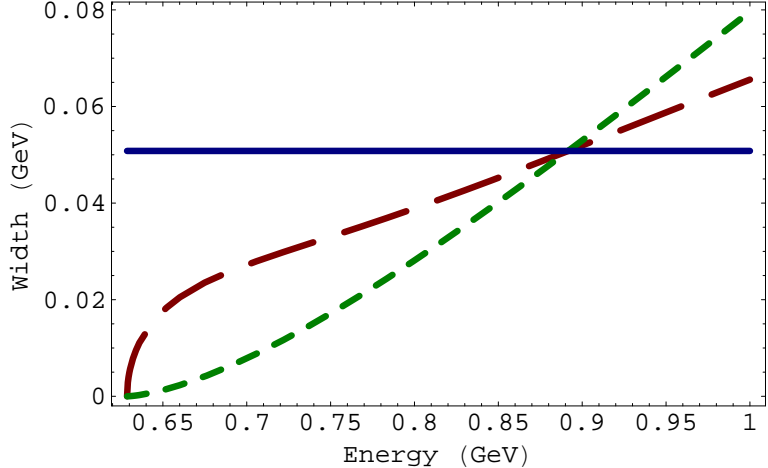}
\caption{\label{widthp} Width behavior.  The solid line (blue) is the experimental value (PDG), the short dashed line (green) is the width function before corrections and the long dashed line (red) is the width function obtained in this work by bringing the propagator into the complex mass form.}
\end{figure}	

\subsection{Width}
The width function can be obtained from Eqn. (\ref{widthf}), by replacing the mass function Eqn. (\ref{massf}). It takes the following form:
\begin{equation}
\bar \Gamma = \frac{\beta q^2+m^2(\gamma-\beta)}{\sqrt{1+(\gamma-\beta)^2} \sqrt{m^2-\beta q^2(\gamma -\beta)}}.
\end{equation}
The experimental value of the width corresponds to evaluate the above expression at $q^2=M_0^2$, which  in turn takes the simple form:
 \begin{equation}
\bar \Gamma_0 =\gamma_0 M_0,
\end{equation}
which is nothing else but the original relation Eqn. (\ref{width0}) between the mass and the width at the mass pole. In addition, since $\beta_{th}=\gamma_{th}=0$, the width function is null at threshold. In Figure \ref{widthp}, we exhibit the corresponding width  behavior and compare with the constant and standard momentum dependence definition Eqn. (\ref{width0}). Note that, although the energy dependent width functions take the same value at threshold and at the pole, they approach these  values in a very different form.

\section{ Longitudinal correction}

The $\tau^-(l) \to K_s(q_1)\pi^-(q_2) \nu(l')$ decay  involves both vector and scalar form factors for its description \cite{zphys,Jamin:2001zq,DescotesGenon:2006uk}. The Belle experimental data \cite{bellekpi} exhibited a small substructure on the $K_s\pi^-$ spectrum around 0.7 GeV that might be explained by a modified scalar form factor  \cite{Moussallam:2007qc,Jamin:2008qg,Boito:2008fq,Boito:2010me,Bernard:2013jxa,Escribano:2014joa}, so far no definite answer has been found. Is in this same scenario where the corrections we have studied might have an effect. Thus, we will obtain the corresponding vector and scalar form factors entering in the description of the process upon the inclusion of the loop corrections discussed in the previous sections. In particular, we expect the longitudinal flavor symmetry breaking correction to modify the scalar form factor. In this section, we follow the standard procedure as in \cite{zphys}. The probability amplitude can be set as:
\be
{\cal M}= V_{us}\frac{G_F}{\sqrt{2}}L_\mu H^\mu
\ee
where $L_\mu\equiv\bar u_\nu\gamma_\mu(1-\gamma_5)u_\tau$ is the leptonic current and $H^\mu$ is the hadronic current, which can be split into the vector and scalar parts as follows:
\be
H^\mu=F_V^{K\pi}(q^2)\left(g^{\mu\nu}-\frac{q^\mu q^\nu}{q^2}\right)(q_1-q_2)_\nu+ F_S^{K\pi}(q^2)q^\mu
\label{hcurrent}
\ee
where $q\equiv q_1+q_2$, and $F_{V,S}^{K\pi}(q^2)$ are the vector ($V$) and scalar ($S$) form factors respectively, which must satisfy:
\be
\text{lim}_{q^2\to 0} q^2 F_S^{K\pi}(q^2)=\Delta^2_{K\pi}F_V^{K\pi}(q^2),
\label{limite0}
\ee 
where $\Delta^2_{K\pi}\equiv(m_K^2-m_\pi^2) $. In addition, the form factors for the decay into $K^-\pi^0$ and $\bar K^0 \pi^-$ are related by isospin symmetry by $F_{V,S}^{K^-\pi^0}=F_{V,S}^{\bar K^0\pi^-}/\sqrt{2}$. 
The form factors, upon a Taylor expansion around $q^2=0$, can be set as follows:
\be
F_V^{\bar K^0\pi^-}(q^2)=f(0)\left(1+\frac{1}{6}<r^2>_V^{K\pi}q^2+...\right)
\ee
\be
F_S^{\bar K^0\pi^-}(q^2)=\frac{\Delta^2_{K\pi}}{q^2}f_S(0)\left(1+\frac{1}{6}<r^2>_S^{K\pi}q^2+...\right)
\label{expansion}
\ee
where $f(0)=f_S(0)=1$ and $<r^2>_{V,S}$ are the vector ($V$) and scalar ($S$) square radius of the meson. On the other hand, the hadronic current can be effectively described assuming that it is dominated by resonances. Usually, considering the $K^*(892)$ and $K^{'*}(1410)$ vectors and either one or two scalars  $\kappa(800)$  and $K_0^*(1430)$ \cite{bellekpi,zphys}.
In the following we consider the case with a single scalar, denoted in general as $K_0^*$. The result can be extended straightforward to two scalars:
 \be
 H^\mu=\left[ g_1D^{\mu\nu}_{K^*}(q^2)+ g_2D^{\mu\nu}_{K^{'*}}(q^2) \right] (q_1-q_2)_\nu
 +g_0\frac{q^\mu}{q^2-m^2_{K^*_0}-i m_{K^*_0} \Gamma_{K^*_0}},
 \ee
 where we have used a short notation for the corresponding strong couplings $g_1\equiv g_{K^*K\pi}$, $g_2\equiv g_{K^{'*}K\pi}$ and $g_0 \equiv g_{K^*_0 K\pi}$ and the propagators, masses and widths carry a subindex to denote the corresponding particle.
  Using the modified propagator Eqn. (\ref{punlazo}), the hadronic current can be set into Eqn. (\ref{hcurrent}) form:
 \bea
 H^\mu&=&\left( \frac{g_1}{BW_{K^*}(q^2)} +\frac {g_2}{BW_{K^{'*}}(q^2)}\right)
 \left( g^{\mu\nu}-\frac{q^\mu q^\nu}{q^2}\right) (q_1-q_2)_\nu+\\
&& 
\left( \frac{\Delta^2_{K\pi}}{q^2}\right)\left( \frac{g_1}{DL_{K^*}(q^2)} +\frac {g_2}{DL_{K^{'*}}(q^2)}  
+\frac{g_0 q^2}{q^2-m^2_{K^*_0}-im_{K^*_0} \Gamma_{K^*_0}}\right)q^\mu\nonumber
 \eea
where $BW(q^2)\equiv m^2-q^2-i Im \Pi^T$ and $DL(q^2)\equiv m^2-i Im\Pi^L $. We can identify the vector and scalar form factors in the above equation \footnote{we have redefined $g_0\equiv g_0(m^2_K-m^2_\pi)$}. Note that they fulfill the relation Eqn. (\ref{limite0}) in the limit $q^2\to 0$. There, they satisfy that $Im \Pi^{L,T}(q^2 \to 0) \to 0$.
A relation between $g_1$ and $g_2$ can be set from the expansion Eqn. (\ref{expansion}) and imposing the restriction $f_S(0)=1$,  such that the scalar form factor becomes:
\bea
F_S^{\bar K^0\pi^-}(q^2)&=&\frac{\Delta^2_{K\pi}}{q^2}
\left[ \frac{1}{(1+\lambda)}
\left(
\frac{m^2_{K^*}}{DL_{K^*}(q^2)} +\lambda \frac {m^2_{K^{'*}}}{
DL_{K^{'*}}(q^2)}
\right) \right. \nonumber\\
&&\left. +\frac{g_0 q^2}{q^2-m^2_{K^*_0}-im_{K^*_0} \Gamma_{K^*_0}}\right]
\eea
where $\lambda$ is in general a complex parameter linked to the original couplings by: $g_1=m_{K^*}^2/(1+\lambda)$ and $g_2=\lambda m_{K^{'*}}^2/(1+\lambda)$. Using the same arguments, 
the vector form factor becomes:
\be
F_V^{\bar K^0\pi^-}(q^2)=\frac{1}{1+\lambda}\left[
 \frac{m^2_{K^*}}{BW_{K^*}(q^2)} +\lambda \frac {m^2_{K^{'*}}}{BW_{K^{'*}}(q^2)}
\right]
\ee 
Note that, using the Taylor expansion and the general functions for the transverse and longitudinal contributions Eqn. (\ref{fun}), we can identify a relation for the vector and scalar radius:
\be
\frac{1}{6}<r^2>_S^{K\pi}\approx \frac{i}{1+\lambda}\left( \frac{ \gamma_1}{ m^2_{K^*}}+ \lambda\frac{\gamma_2}{m^2_{K^{'*}}}-g_0\frac{ \Gamma_{K^*_0}}{m^3_{K^*_0}}\right)- \frac{g_0}{m^3_{K^*_0}}
\ee
and
\be
\frac{1}{6}<r^2>_V^{K\pi}\approx \frac{1}{1+\lambda}
\left( \frac{1}{ m^2_{K^*}}+ \frac{\lambda}{m^2_{K^{'*}}} +i\left[
\frac{ \beta_1}{ m^2_{K^*}}+ \lambda\frac{\beta_2}{m^2_{K^{'*}}}\right]\right),
\ee
where the subindex $i=1$ and 2 correspond to $K^*$ and $K^{'*}$  respectively. 
Using the above results, we can solve the equations consistently to determine the $\lambda$ and $g_0$ parameters, for a given set of values of the vector and scalar radius. They can also be used as free parameters in the fit to the $\tau \to K\pi \nu$ spectrum. This last is the procedure we will follow.

We can obtain the corresponding description for the form factors for the complex mass form of the propagator by noticing that, in this case, the transverse and longitudinal corrections are taken as equal  $ Im\Pi^L=Im\Pi^T=\Gamma(q^2)\sqrt{q^2}$. Provided the mass and width are functions of the energy and the mass and width parameters as defined in Eqs. (\ref{massf}) and (\ref{widthf}).

\subsection{$K \pi$ spectrum}
\begin{figure}
\includegraphics[width=4in, angle=-90]{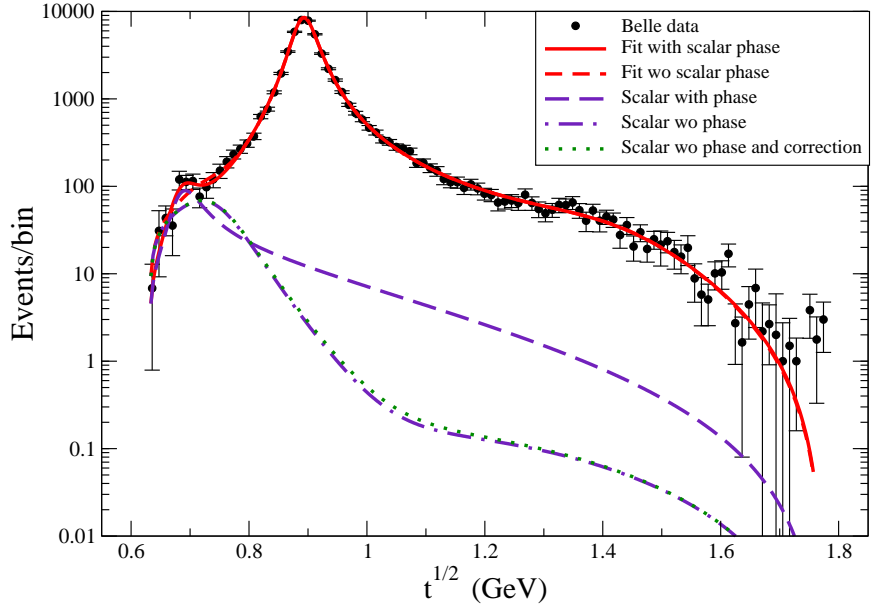}
\caption{\label{fit2v1s} The $K \pi$ spectrum for the $\tau\to K_S\pi^-\nu_\tau$ decay. Symbols are the data from Belle experiment, the solid line is the fit considering the vector and scalar form factors, the dashed line is the scalar contribution from the fit.}
\end{figure}

The $K \pi$ spectrum for the $\tau\to K_S\pi^-\nu_\tau$ decay is given by:
\bea
\frac{d\Gamma(\tau\to K_S\pi^-\nu_\tau)}{d\sqrt{t}}&=&\frac{G_F^2|V_{us}|^2m_\tau^3 S_{EW}}{96\pi^3t}
\left(1-\frac{t}{m_\tau^2}\right)^2\nonumber\\
&&\left[\left(1+2\frac{t}{m_\tau^2}\right) q^3_{K\pi}|F_V^{K\pi}(t)|^2+
\frac{3 \Delta^2_{K\pi}q_{K\pi}}{4t}|F_S^{K\pi}(t)|^2
\right],
\eea
where $t\equiv (q_1+q_2)^2$, $q_{K\pi}= \sqrt{t^2-2t\Sigma_{K\pi}+\Delta^2_{K\pi}}/(2\sqrt{t})$, $\Sigma_{K\pi}\equiv(m_K^2+m_\pi^2)$, $S_{EW}=1.0201$ and $V_{us}=0.2163$.
Using the form factors obtained above, we fit the experimental data from Belle \cite{bellekpi} considering a single scalar, with and without a phase ($g_0$ taken as complex or real respectively). In Figure \ref{fit2v1s}, we exhibit the fit with  scalar phase (solid line) and without phase (short-dashed line). The only visible difference is found in the low energy region where the freedom on the phase allows a better description of the data exhibiting a bump around 0.7 GeV. The scalar contribution with and without phase are represented by the dashed and dot-dashed lines respectively. The longitudinal correction from the vector particles obtained in this work is only visible in the latter case (dotted line) in the region around 1 GeV. The fitted parameters are shown in Table \ref{fitpar1}. The inclusion of a second scalar ($K^*_0(1430)$) fades out the features exhibited by a single scalar, we have shown this case for illustrative purposes, since it is the only case were the longitudinal correction can be identified out of the other contributions.

\begin{center}
\begin{table}
\begin{tabular}{|c|c|c|}
\hline
Parameter & without phase& with phase\\
\hline      $m_{K^*}  $    &   0.8953   $\pm$    0.0002 & 0.8953     $\pm$   0.0002\\
      $\Gamma_{K^*}   $  &   0.0475 $\pm$  0.0005 &  0.0475  $\pm$   0.0005\\
      $m_{K^{'*}}$         & 1.41    $\pm$   0.04 &  1.41      $\pm$   0.04\\
       $\Gamma_{K^{'*}}$  &  0.42   $\pm$    0.03 &  0.43     $\pm$   0.03\\
      $m_{K^*_0} $      &  0.75    $\pm$   0.02  & 0.69     $\pm$   0.01\\
      $\Gamma_{K^*_0}$   &  0.15    $\pm$   0.04 & 0.06  $\pm$  0.02\\  
      $|\lambda|$      &  -0.45    $\pm$   0.06  &  -0.47     $\pm$   0.06 \\
      $\phi_\lambda$       &  0.8    $\pm$   0.03 &  0.82     $\pm$   0.02\\
      $|g_0|$        & -0.5    $\pm$   0.1   &  -0.4      $\pm$  0.1 \\ 
      $\phi_{g_0}$ & - & 2.1      $\pm$   0.5\\
$\chi^2/dof$&92/99 & 89/99 \\
\hline  
\end{tabular}
\caption{Parameters from the fit to $K\pi$ spectrum. A single scalar resonance case. Masses and widths are in GeV.}
\label{fitpar1}
\end{table}
\end{center}

\section{Conclusions}	
We have shown that a non trivial modification to the propagator of vector mesons, induced by including the absorptive corrections, can be set into a complex mass form, provided the mass and the width parameters are replaced by energy dependent functions fulfilling the same requirements at the pole. We have considered the case for the $K^*(892)$ vector meson, to illustrate the meaning of the parameters involved. In particular, we have found that the mass function set to  match the complex mass form  deviates around 2 MeV from the $K\pi$ threshold to the pole position. This may contribute to explain the mass difference observed between the neutral and charged $K^*$ \cite{pdg}, since the neutral particle is not influenced by these corrections. On the other hand, the width energy dependence is modified exhibiting a different evolution between its value at threshold and at the pole. We can trace back the main source of the modification to the longitudinal correction. By setting this to zero the modification becomes negligible. We would like to stress that, in order to be consistent with the gauge invariance, the use of a complex mass propagator must employ the mass and width functions obtained in this work, which  may be important in precision measurements.

We have exemplified the role of the corrections, at the level of the form factors, considering the $\tau \to K_s\pi^-\nu$ decay. The modified vector is found to be the same as in the complex mass form, while the scalar form factor receives a modification from the longitudinal correction to the vector propagator. The $K\pi$ spectrum was fitted considering three resonances. Nonetheless, for this particular case, the induced correction to the scalar form factor was found to be negligible. Another scenarios with stronger flavor symmetry breaking, as those involving $D^*$ mesons,  and/or scalar dominated processes might be more sensitive to this kind of corrections.

We thank Dr. P. Roig for a careful reading of the manuscript and very useful suggestions.

\end{document}